\documentstyle[aps,prl,epsf]{revtex}
\begin{document} 
 
\draft 
 
\title{ 
Symmetries and Fixed Point Stability of Stochastic Differential 
Equations Modeling Self-Organized Criticality 
} 
 
\author{\'Alvaro Corral\cite{email.alvaro} and Albert 
D\'{\i}az-Guilera\cite{email.albert}}  
 
\address{ 
Departament de F\'{\i}sica Fonamental \\ Facultat de F\'{\i}sica, 
Universitat de Barcelona \\ Diagonal 647, E-08028 Barcelona, Spain } 
 
\date{\today} 
 
\maketitle 
 
\begin{abstract} 
 
A stochastic nonlinear partial differential equation is built for
two different models exhibiting self-organized criticality,
the Bak, Tang, and Wiesenfeld (BTW) sandpile model and the Zhang's model.
The dynamic renormalization group (DRG) enables to compute 
the critical exponents.
However, the nontrivial stable fixed point of the DRG transformation
is unreachable for the original parameters of the models.
We introduce an alternative regularization of the 
step function involved in the threshold condition, which breaks the 
symmetry of the BTW model.
Although the symmetry properties of the two models are different,
it is shown that they both belong to the same universality class. 
In this case the DRG procedure leads to a symmetric behavior for both 
models, restoring the broken symmetry,
and makes accessible the nontrivial fixed point. 
This technique could also be applied to other problems with
threshold dynamics. 
\end{abstract} 
 
 
\section{Introduction} 
 
In the last decade much attention has been paid to the phenomenon known 
as self-organized criticality (SOC). Bak, Tang, and Wiesenfeld (BTW) 
\cite{pra38.364} studied a cellular automaton model as a paradigm for 
the explanation of two widely occurring phenomena in nature: $1/f$ 
noise and fractal structures.  Both have in common a lack of 
characteristic scales.  
The SOC models, although not always show $1/f$ noise,
have no characteristic scales too,  
and this scale invariance suggests that these 
systems are critical in analogy with classical equilibrium critical 
phenomena, but in SOC one deals with dynamical 
nonequilibrium statistical properties. Moreover, the system 
evolves naturally to the critical state without any tuning of 
external parameters, that is, in a self-organized process. 
 
Several cellular automata and coupled map lattices 
models exhibiting SOC 
have been reported in the literature. In the original 
sandpile model of Bak {\em et al.} \cite{pra38.364} the system is 
perturbed externally by a random addition of sand grains. Once the 
slope between neighboring cells has reached a threshold 
value, sand is transferred between them in a fixed amount. 
Taking this model as a reference, different dynamical rules have been 
investigated leading to a wide variety of universality classes.
Continuous variables with a full transfer from a cell instead of a 
fixed discrete amount 
\cite{prl63.470,p173a.22,pra44.1386,pra45.8551},  
directed flows \cite{pra39.6524}, threshold condition imposed on the 
height, on the gradient, or even on the laplacian \cite{p179a.249}, 
anisotropy \cite{pra42.769}, are a few examples. 
These randomly driven models do not exhibit SOC when the 
interaction rules are not conservative \cite{jsp22.923}. Later 
on, other deterministically driven models have been introduced 
where conservation is not a necessary condition 
\cite{prl66.2669,prl68.1244,prl68.2417,pra46.r1720,prl1,review}. 
Much more recently, sandpile models with deterministic perturbations
but intrinsic randomness in the threshold dynamics
have been used to reproduce experimental results on transport
properties on rice piles \cite{Christensen96}.
The close connection between these sandpile models and interface depinning
has been establish in Ref. \cite{Paczuski96}.
 
Some authors have attempted to connect 
the randomly driven models to stochastic 
differential equations 
\cite{prl62.1813,prl64.1927}. These continuous 
descriptions are built according to the symmetry rules obeyed by the 
discrete models in order to achieve a generic scale invariance 
condition \cite{nato}.  
Nevertheless none of them neither explicitly nor implicitly 
includes the threshold condition,
which is one of the main characteristics of SOC models.
On the other hand anomalous diffusion equations with 
singularities in the diffusion coefficient have been considered in 
order to study the deterministic dynamics of the avalanches generated 
in the critical state 
\cite{prl65.2547,prl68.2058}.
Latterly, a different approach has been introduced by Pietronero  
and co-workers, using a real-space renormalization procedure
to determine the dynamical exponent as well as the avalanche
size exponent. 
\cite{Pietronero}.
 
        In a previous paper \cite{el26.177} 
one of us studied two nonlinear stochastic differential 
equations derived from the discrete dynamical rules of two models 
with different symmetry properties. In principle one would 
expect, due to this reason,  
different critical behavior. 
However, it was shown 
analytically, by means of the dynamic renormalization group (DRG) 
\cite{pra16.732,pra39.3053,Barabasi}, 
that both models belong to the same universality class.
The threshold condition was kept but the 
step function was regularized in order to allow a power series 
expansion. 
In the limit that recovers the threshold
it was shown that the coupling constants that distinguish both models
become decoupled of the common coupling constants;
since the critical exponents only depend on the latter constants,
one obtains the same values for both models.
Once this equivalence was established, 
the most symmetric model was considered, 
showing that an infinite number of coupling constants was 
relevant below the upper critical dimension $d_c=4$; by expanding in 
the number of nonlinearities, the DRG procedure, 
up to first order in $\epsilon=4-d$, gave 
an estimate of the dynamical exponent close to the value 
obtained by scaling arguments and  
in the numerical simulations \cite{prl63.470,pra45.8551}.
This value of the dynamical exponent is obtained when the flow in 
parameter space reaches the nontrivial stable fixed point;
nevertheless, when taking into account the physical values of the 
parameters, they do not lie in the basin of attraction of the fixed 
point, thus making this computed value in some sense speculative,
since it cannot be ensured that the flow in parameter space will be 
able to reach the  attractor. 

     Our goal in this paper is to complement the previous work in 
order to check the validity in the calculation of the critical 
exponents at the stable fixed points and to analyze the role played 
by symmetries in randomly driven SOC models
and, in general, in other models where an infinite hierarchy of 
nonlinear terms is required.
Our procedure also illustrates the effect of symmetry
breaking in DRG calculations
as a mechanism to make the attractors in parameter space
accessible for the physical values of the parameters in the 
original equations.
The continuum equation for the BTW and Zhang's models 
is introduced in Sect. II, as well as the alternative regularization 
which breaks the symmetry that distinguish
both models.
In Sect. III we develop the DRG procedure and show that 
this symmetry is irrelevant,
in view of the fact that the nontrivial fixed point is not
modified by this new approach.
Moreover, the effect of symmetry breaking allows the flow of the original 
parameters to reach the nontrivial fixed point,
where the critical exponents can now be computed.
Finally we present our conclusions in Sect. IV. 
 
\section{Models and Symmetries} 
 
        First, we describe briefly the dynamics of the two  
SOC models under consideration. 
The first model was originally proposed by Zhang 
\cite{prl63.470}, and 
it consists of a $d$-dimensional lattice in which  
any site can store some continuously distributed variable $E$. 
This variable, that we will call energy, 
can have different physical interpretations \cite{p173a.22}.  
The system is perturbed by adding a random amount of 
energy $ \delta E >0 $ at a randomly chosen site.  
Once a site reaches a 
value of the energy greater than the threshold value $E_c$, this site 
becomes active, and transfers all its energy 
to its nearest neighbors.  
At this point the input of energy from the outside is turned off.  
The energy transferred to the neighboring sites can make them active, 
giving rise to an activation cluster or avalanche, that ends when all 
the sites have reached a value of the energy smaller than $E_c$. 
It is only when the avalanche has stopped that energy is added again, 
otherwise the system remains quiescent.   
In this way there is a clear time-scale separation in the dynamics. 
The external noise acts in a slow time scale, whereas the avalanches 
evolve infinitely fast, in comparison. 
The second model differs from Zhang's model only in the amount of  
energy an active site transfers to its neighbors, which is a fixed  
amount $E_c$, instead of its whole energy $E$. 
Therefore it is closer to the original sandpile model of Bak  
{\em et al.} \cite{pra38.364}, but continuous in $E$. 
When $\delta E$ is not random but fixed 
this difference becomes irrelevant. 
For this reason, it will be referred as BTW model. 
Notice that both models are conservative, in the sense that the 
added energy (always positive) is only dissipated  
at the open boundaries. 
 
The microscopic evolution rules can be written 
from time $t$ (in the fast time scale) to $t+1$ for each site $i$ as 
\begin{mathletters} 
\label{discrete} 
\begin{equation} 
     E_i(t+1)=E_i(t)-E_i(t)\Theta\left( E_i(t)-E_c\right)+ 
     \frac{1}{q} \sum_{nn} E_{nn}(t)\Theta\left(E_{nn}(t)-E_c\right)  
     + \xi _i(t) 
\label{discretezhang} 
\end{equation} 
and 
\begin{equation} 
       E_i(t+1)=E_i(t)-E_c\Theta\left(E_i(t)-E_c\right)+ 
       \frac{1}{q} \sum_{nn} E_c\Theta\left(E_{nn}(t)-E_c\right)  
       + \xi _i(t), 
\label{discretebtw} 
\end{equation} 
\end{mathletters}%
for Zhang's and BTW model, respectively. The sum runs over the $q$ 
nearest neighbors of site $i$, labeled as $nn$, and 
the threshold condition enters through the Heaviside step function  
$ \Theta $, defined as $ \Theta (x<0)=0 $ and $ \Theta (x>0)=1 $. 
Due to the continuous nature of the models the value $ \Theta (x=0) $ 
is irrelevant and we can keep it undefined, by now. 
For the external noise $\xi _i(t)$, 
that drives the system only when there are no active sites, 
one can formally write 
\begin{equation}  
      \xi_i(t)= \delta E\, \delta_{i,n(t)} \prod_{\forall j} 
      \ \left[1-\Theta\left(E_j(t)-E_c\right)\right], 
\label{discretenoise}  
\end{equation} 
where $\delta_{i,n(t)}$ is the Kronecker delta symbol 
and $n(t)$ is a random vector pointing the site of the lattice 
that will be perturbed with a random amount of energy $\delta E$ 
(in the original BTW model $\delta E=E_c/q$). 
The product runs over all the lattice sites. 
 
When applying the DRG one deals with infinite systems and then 
the important effect of dissipation at the open boundaries is not 
taken into account. 
However in SOC models a distribution of absorbing
defects through the lattice plays the same role as the open (absorbing)
boundaries \cite{Maslov}, 
as we have verified through computers simulations \cite{unp}.
Then, we can redefine our models in an infinite lattice but with a quenched
distribution of defects.
The results are not modified with this assumption.
Another possibility is to consider that each site of an infinite lattice
has a small probability of dissipating an amount of energy $E_c/q$
when it topples, instead of transferring it to a certain neighbor.
This procedure, that represents the assumption of random boundaries, 
implies that
when a site receives
a toppling from some neighbor, it has a small probability of not to
accept the amount of energy $E_c/q$, that is lost \cite{Christensen_private}.
This dissipation can be included as a new term in the noise,
and Eq. (\ref{discretenoise}) has to be replaced by  
\begin{equation}  
      \xi_i(t)= \delta E\, \delta_{i,n(t)} \prod_{\forall j} 
      \ \left[1-\Theta\left(E_j(t)-E_c\right)\right]
      \ - \ \sum_{nn} \zeta_{nn} \Theta(E_{nn}(t)-E_c), 
\label{discretenoise2}  
\end{equation} 
where $\zeta_{nn}$ is a dichotomous noise,
taking value 0 with large probability and value $E_c/q$ (dissipation)
with a small one.
If $\zeta_{nn}$ depends on $t$, i.e., $\zeta_{nn}=\zeta_{nn}(t)$
we are dealing with annealed random boundaries, 
whereas if it only depends on the position, 
we have quenched random boundaries or absorbing defects.

In terms of a rescaled energy $E-E_c\rightarrow E$, and introducing 
a parameter $Z$ to unify the description, we have for both models
\begin{equation} 
       E_i(t+1)-E_i(t)= 
       \frac{1}{q} \sum_{nn} \left \{ \left[Z E_{nn}(t)+E_c\right] 
       \Theta\left(E_{nn}(t)\right) 
       -\left[Z E_i(t)+E_c\right]\Theta\left(E_i(t)\right)\right\}  
       + \xi _i(t), 
\label{discreteboth} 
\end{equation} 
where $Z=1$ for Zhang's model and $Z=0$ for BTW. 
Equation (\ref{discreteboth}) 
defines a stochastic coupled map lattice. 
Moreover, notice that the deterministic BTW equation 
displays invariance under a parity transformation of the order  
parameter, $E \rightarrow -E$. 
This one is the only symmetry that the BTW  
model does not share 
with Zhang's model. The common symmetries are invariance under  
spatial translations, 
rotations, and reflections, as well as conservation of the order  
parameter. 
 
Equation (\ref{discreteboth}) can be coarse-grained in order to  
obtain a continuum equation for the effective $E(\vec{r},t)$. 
Then, by using the prescriptions for the temporal derivative and 
for the Laplace operator: 
\begin{equation} 
      \frac{\partial E(\vec{r},t)}{\partial t}= 
      \alpha \nabla ^2 \left \{[Z E(\vec{r},t) + E_c] 
      \Theta(E(\vec{r},t))\right \} 
      +\eta(\vec{r},t), 
\label{continuumexact} 
\end{equation} 
where $\alpha$ is a coefficient that depends on the lattice spacing, 
the unit time step and the coordination number $q$. The noise  
$\eta(\vec{r},t)$ accounts for the effective external  
noise as well as for the internal noise that appears due to the  
elimination of microscopic degrees of freedom. 
 
Up to now, Eq. (\ref{continuumexact}) describes truly the coarse-grained
evolution of the system, but we have not yet characterized the noise 
$\eta(\vec{r},t)$, which derives from $\xi_i(t)$.
The product in Eq. (\ref{discretenoise2}) makes $\eta(\vec{r},t)$
to be a multiplicative noise
that depends on the whole lattice state, 
and the problem is intractable.
We are going to ignore the restrictions imposed by the step
functions in Eq. (\ref{continuumexact}),
breaking the time scale separation.
Then the noise $\eta(\vec{r},t)$ acts continuously in time and can
provoke avalanches to overlap.
However for small enough noise this is very unlikely, and 
one can still identify avalanches in computer simulations.
Moreover the dynamical exponent does not change with this assumption
\cite{pra45.8551},
due to the fact that the added noise is orders of magnitude smaller
than the energy transferred by the avalanche and then its
dynamics is not affected.
In this case we still have two time scales, although they overlap.
In what follows $\eta(\vec{r},t)$ will be considered as an additive
random process,  including
two effects, the external driving, always positive, and the dissipation
at the (random) boundaries, always negative.
In the statistical stationary state the random input of energy must 
equal on average the output at the boundaries.  
Then we assume that
\begin{equation} 
       < \eta (\vec{r},t)> = 0.
\label{noise0} 
\end{equation} 
In fact this is the same assumption done in all the studies of SOC
by means of DRG \cite{prl62.1813,prl64.1927},
and it is somehow equivalent
to the stationary condition used in Ref. \cite{Pietronero}.

Moreover we are mainly interested in the spatio-temporal propagation
of a perturbation through the system, that is, in measuring the value
of the dynamical exponent. 
For this purpose we have to look at the system
in the fast time scale, i.e., the scale of the evolution of the avalanches.
In Ref. \cite{pra45.8551} it was argued that in this case one can understand
the noise as a quenched Gaussian process uncorrelated in space, 
and then its correlation function is given by
\begin{equation} 
       < \eta (\vec{r},t) \eta (\vec{r'},t')> = 2 \Gamma 
       \delta ^d(\vec{r}-\vec{r'}).
       \label{external} 
\end{equation} 
When looking at the system at the slow time scale one cannot use this 
prescription for the noise, which has to be uncorrelated in time too,
i.e., 
$< \eta (\vec{r},t) \eta (\vec{r'},t')> = 2 \Gamma 
       \delta ^d(\vec{r}-\vec{r'})\delta (t-t')$,
and this prescription is mainly related to the interface roughness 
between avalanches \cite{fractals}. 
 
Equations (\ref{continuumexact}), (\ref{noise0}), and (\ref{external}),
together with the fact that the noise is a Gaussian process,
define completely our model.
However, the presence of the step function in Eq. (\ref{continuumexact}) 
gives rise to a strong nonlinearity. A perturbative expansion 
of this equation can be performed if one regularizes the step function 
as 
\begin{equation} 
      \Theta (E) = \lim_{ \beta  \rightarrow  \infty } f(\beta E), 
\label{theta} 
\end{equation} 
and makes a series expansion of $f(\beta E)$ in powers of $E$ 
\cite{prl68.2058,el26.177}. 
The function $f(x)$ must be monotonously increasing with 
$f(-\infty)=0$ and $f(\infty)=1$. 
Moreover we choose $f(x)-1/2$ as an odd function,
so $f(0)=1/2$. 
Several functions of this type have been used in the literature, 
but that coming from the error function as 
\begin{equation} 
      f(x)= \frac{1+erf(x)}{2}=\frac{1}{\sqrt{\pi}} 
      \int_{-\infty}^{x}e^{-y^2}dy, 
\label{error} 
\end{equation} 
allows a power expansion that has an infinite radius of convergence,
in contrast with previous choices \cite{prl68.2058,el26.177}. 
In any case, the relevant results do not depend on the particular form of
$f(x)$.

The regularization given by Eq. (\ref{theta}) 
keeps the symmetry of the step function  
and therefore the invariance under a parity transformation in the  
BTW model. 
As an alternative regularization that breaks  
this invariance we propose the following: 
\begin{equation} 
      \Theta (E) = \lim_{ \beta  \rightarrow  \infty } f(\beta E + K) 
\label{alternative} 
\end{equation} 
with $K$ an arbitrary constant.  
Although in the limit $\beta \rightarrow \infty$ we recover the step 
function, we do not recover its symmetry anymore, because 
$\Theta(E=0) = f(K) \ne 1/2$ if $K \ne 0$, and 
this is the reason for the breaking of the symmetry in the BTW model. 
Now we perform a series expansion of the regularizing  
function $f(\beta E)$ in powers of $E$, obtaining
\begin{equation} 
    \Theta (E) =  
    \lim_{ \beta  \rightarrow  \infty } \sum_{n=0}^{\infty} a_n ( 
    \beta,K) E^n, 
\label{expansion} 
\end{equation} 
where the coefficients $a_n(\beta,K)$ are given by 
\begin{equation} 
       a_n(\beta,K)=\frac{f^{(n)}(K) \beta^n}{n!}, 
\label{an} 
\end{equation} 
being $f^{(n)}(K)$ the $n$-th order derivative of $f(x)$ at $x=K$. 
Substituting the expansion (\ref{expansion}) into  
Eq. (\ref{continuumexact}) we can write 
\begin{equation} 
      \frac{ \partial E(\vec{r},t)}{ \partial t}=D \nabla 
      ^{2}E(\vec{r},t)+ \sum_{n=2}^ \infty   \lambda _n  \nabla 
      ^{2}E^n(\vec{r},t)+ \eta (\vec{r},t), 
\label{series} 
\end{equation} 
where the effective diffusion constant $D$ and the coupling 
constants $ \lambda _n$ (that make the equation nonlinear) 
take different values depending on the model: 
\begin{eqnarray} 
       D & = &   \lim_{\beta \rightarrow \infty}\alpha  
       (E_c f^{(1)}(K)\beta  +  Z f(K)), 
\label{D} 
\\ 
      \lambda _n  &  =  &  
      \lim_{\beta \rightarrow \infty} \frac{\alpha \beta^n}{n!} 
      \left(E_c f^{(n)}(K) + Z \frac{\ n f^{(n-1)}(K)}{\beta}
      \right), 
      \hspace{2em} 
       n = 2, 3,...,\infty. 
\label{lambdan} 
\end{eqnarray} 
On the one hand, for $K=0$, since all the even derivatives verify 
$f^{(2n+2)}(0)=0$, all even coupling constants vanish for the 
BTW model, whereas they do not for the Zhang's one. 
Using Eq. (\ref{series}) this fact allows to verify the symmetry 
of the BTW model under the parity transformation of the order  
parameter $E$. 
On the other hand, for $K \ne 0$, the even coupling constants do not 
vanish in any case, and this fact constitutes the symmetry breaking 
for the BTW model.  
Then, under this condition, the only difference between both models 
is that the 
constants depend on $ \beta $ in a different way; however 
it is easy to see that in the limit $ \beta  \rightarrow   
\infty $ both sets of 
constants are identical 
and then the Zhang's model and the broken-symmetry 
BTW model have to belong to the same universality class. 
This fact can only be shown for $K \neq 0$. 
Nevertheless, the fact of considering $K = 0$ only introduces a 
difference  
in the value of $\Theta(0)$, which is irrelevant in a continuous 
model, and then one can include the (symmetric) BTW model in this 
universality class too.

At this point it is worth noting that we have transformed a 
stochastic coupled map lattice, that involves a threshold 
condition and presents a clear separation of time scales, into a 
nonlinear stochastic partial differential equation, where the 
nonlinearity of the threshold 
is described by an infinite series of powers and 
the randomness enters via a Gaussian process, with zero mean to 
account for the dissipation at the boundaries.  
Along this 
transformation, and due to the approximations we have 
performed concerning the noise correlation,
we have broken the time 
scale separation since the noise acts constantly in time.
Nevertheless, we expect that 
such an equation explains the dynamical properties of the system 
within the fast time scale of the propagation of the avalanches. 
As we have mention before and 
it is discussed in \cite{fractals}, to deal with the slow 
time scale, where the avalanches are instantaneous, another 
noise correlation is more appropriate. 
 
\section{Dynamic Renormalization Group Procedure} 
 
The model to be studied by the DRG is defined by the         
nonlinear partial differential equation (\ref{series})
and the Gaussian noise given by (\ref{noise0}) and (\ref{external}).  
As a previous step we can check the relevance of the different 
coupling constants in this equation by naive dimensional analysis: a 
change of scale $ b=e^l>1$:
\begin{equation} 
    \vec{r'}= e^{-l} \vec{r},\hspace{3em} 
    t' = e^{-zl}t,\hspace{3em} 
    E'=  e^{ -\chi l}E, 
\label{scale} 
\end{equation} 
is performed in Eqs. (\ref{series}) and (\ref{external}), 
being $\chi$ the roughness 
exponent, which is related to the hydrodynamic exponent,  
and $z$ the dynamical exponent. 
Then one obtains that the parameters transform as
\begin{equation}
    D \rightarrow b ^{z-2}D, \hspace{3em} 
    \Gamma \rightarrow b ^{2(z-\chi)-d} \Gamma, \hspace{3em} 
    \lambda_n \rightarrow b ^{z+(n-1)\chi-2}\lambda_n.
\end{equation} 
Under this scaling 
transformation, $z$ and $ \chi $ are chosen to keep the linear 
model scale invariant, i.e., the parameters $D$ and $\Gamma$
have not to be modified. This choice gives $z=2$, $\chi=(4-d)/2$, and
\begin{equation}
    \lambda_n \rightarrow b ^{\frac{4-d}{2}(n-1)}\lambda_n.
\end{equation} 
Then one can see that 
when we apply iteratively the transformation ($b \rightarrow
\infty $) for $d>4$ all the nonlinear terms vanish and they are  
irrelevant.
However,
all the coupling constants go to infinity 
for $d < 4$, and hence all nonlinear terms become relevant;
this implies that the upper critical dimension is $d_c=4$,
and nontrivial values of the exponents are expected below it.
 
The relevance of all the terms makes our problem
much more complicated than for instance the 
Kardar-Parisi-Zhang model of interface growth,
where only the first nonlinear term is relevant \cite{kpz}.
The appropriate treatment of Eq. (\ref{series})
would be to renormalize the infinite number of relevant
coupling constants that are involved.
Of course this is impossible to do in practice.
In \cite{el26.177} an expansion in the number of coupling constants 
for the BTW model was performed with only odd terms,
i.e., without symmetry breaking ($K=0$).
The critical exponents where obtained as a function of the
highest coupling constant, up to $\lambda_9$.
Fortunately, the dynamical exponent was well behaved
and could be extrapolated up to $\lambda_{\infty}$.
However, keeping the symmetry of the step function, 
the nontrivial fixed point of the
DRG is unreachable using the parameters given by Eqs. 
(\ref{D}) and (\ref{lambdan}), even for the Zhang's model.
We want to show that with the proposed alternative regularization
of the step function, that breaks the symmetry of the BTW model
and allows the existence of even coupling constants, the DRG fixed
points are not changed but now they are accessible to the flow 
when the parameters take their real values.
For this reason, and as a first attempt to justify our hypothesis
as well as the conclusions of Ref. \cite{el26.177}, we will focus
on Eq. (\ref{series}) with only its first two nonlinear terms,
i.e., $ \lambda _2$  
and $ \lambda _3$, and see how they 
behave under a DRG transformation, 
\begin{equation} 
 \frac{ \partial E(\vec{r},t)}{ \partial t}=D \nabla ^{2}E(\vec{r},t)+ 
 \lambda _2  \nabla ^{2}E^2(\vec{r},t)+ \lambda _3   
 \nabla ^{2}E^3(\vec{r},t)+ \eta (\vec{r},t). 
\label{order3} 
\end{equation} 

    The DRG procedure consists of the removal 
of the fast modes (large wavenumber $k$) in the momentum space, 
followed 
by a rescaling of a factor $e^l$ in order to recover the original 
Brillouin zone \cite{pra16.732,pra39.3053,Barabasi}. 
After this transformation, one obtains an equation 
that is equivalent to the original one 
but with different (effective or renormalized) coefficients.  
Successive iterations of this  
transformation give the flow of the coefficients in the parameter 
space. 
If this flow converges towards a fixed point, the system presents 
''scale invariance'' in the hydrodynamic limit (large-distance and  
long-time behavior). 
Then, the fluctuations of the order parameter verify the scaling 
equation 
\begin{equation} 
      <\left[ E(\vec{r}_0,t_0) - E(\vec{r}_0+ 
      \vec{r},t_0+t) \right]^{2} >^{ \frac{1}{2} } \sim  r^{ \chi } 
      F(t/r^z), 
\label{scaling} 
\end{equation} 
where the critical exponents  $\chi $ and $z$  
are those that ensure the existence of the fixed point.
However, it is worth mentioning that with this procedure 
the scaling function $F(x)$ remains unknown \cite{Hwa-Frey}. 
 
We now outline the DRG calculation. 
First of all we write Eq. (\ref{order3}) in Fourier space: 
\begin{equation} 
     E= G_0\eta  
     - G_0 \frac{\lambda_2 k^2}{(2\pi)^{d+1}}E*E 
     - G_0 \frac{\lambda_3 k^2}{(2\pi)^{2(d+1)}}E*E*E. 
\label{fourier_equation} 
\end{equation} 
Here $E(\vec{k},\omega)$ and $\eta(\vec{k},\omega)$ are defined as 
the Fourier transforms of $E(\vec{r},t)$ and  
$\eta(\vec{r},t)$, i.e.,
\begin{equation} 
      E(\vec{k},\omega)= 
      \int d^d r\ dt\ e^{i(\omega t-\vec k \cdot \vec r)}E(\vec{r},t), 
\label{fourier_definition} 
\end{equation} 
whereas 
\begin{equation} 
       G_0(k,\omega)=\frac{1}{-i \omega + D k^2} 
\label{propagator} 
\end{equation} 
is called the bare propagator. The symbol ''$*$'' represents the 
convolution product, defined as 
\begin{equation} 
      (E*E)(\vec{k},\omega)= 
      \int d^d q\ d\Omega \  
      E(\vec q,\Omega)E(\vec k-\vec q,\omega-\Omega).      
\label{convolution} 
\end{equation} 
Fig. 1 (a) shows the expression of 
Eq. (\ref{fourier_equation}) in terms of Feynman diagrams. 
As the intensity of the noise $\Gamma$ 
has also to be renormalized by the DRG transformation, 
we need to consider the equation for the correlation function of the 
transformed energy $<E(\vec{k},\omega)E(\vec{k'},\omega')>$,  
which, up to one-loop order, is: 
\begin{equation} 
  <EE'>=G_0G_0'<\eta\eta'> 
  +\frac{\lambda_2^2 k^2 {k'}^2 G_0G_0'} 
  {(2\pi)^{2(d+1)}}<(E*E)(E*E)'>, 
\label{correlationE} 
\end{equation} 
where the prime denotes a dependence on $\vec k ', \omega '$ instead 
of the dependence on $\vec k , \omega $. 
The diagrammatic  
representation of this equation is shown in Fig. 1(b).  
Eqs. (\ref{fourier_equation}) and (\ref{correlationE}),
that are the ones that we are going to renormalize, 
hold for $0<k<\Lambda$, where $\Lambda$ is the wavenumber cutoff  
due to the underlying discrete structure. 
The transformed noise $\eta(\vec{k},\omega)$ turns out to be 
also a Gaussian process with zero mean, but with correlation 
\begin{equation} 
       < \eta (\vec{k},\omega)\eta (\vec{k}',\omega')> = 
       2 (2\pi)^{d+2}\Gamma 
       \delta ^d(\vec{k}+\vec{k}') \delta(\omega) \delta(\omega'). 
\label{fourier_noise} 
\end{equation} 
\begin{figure}
[htbp] 
\epsfxsize=5truein 
\hskip 0.15truein
\epsffile{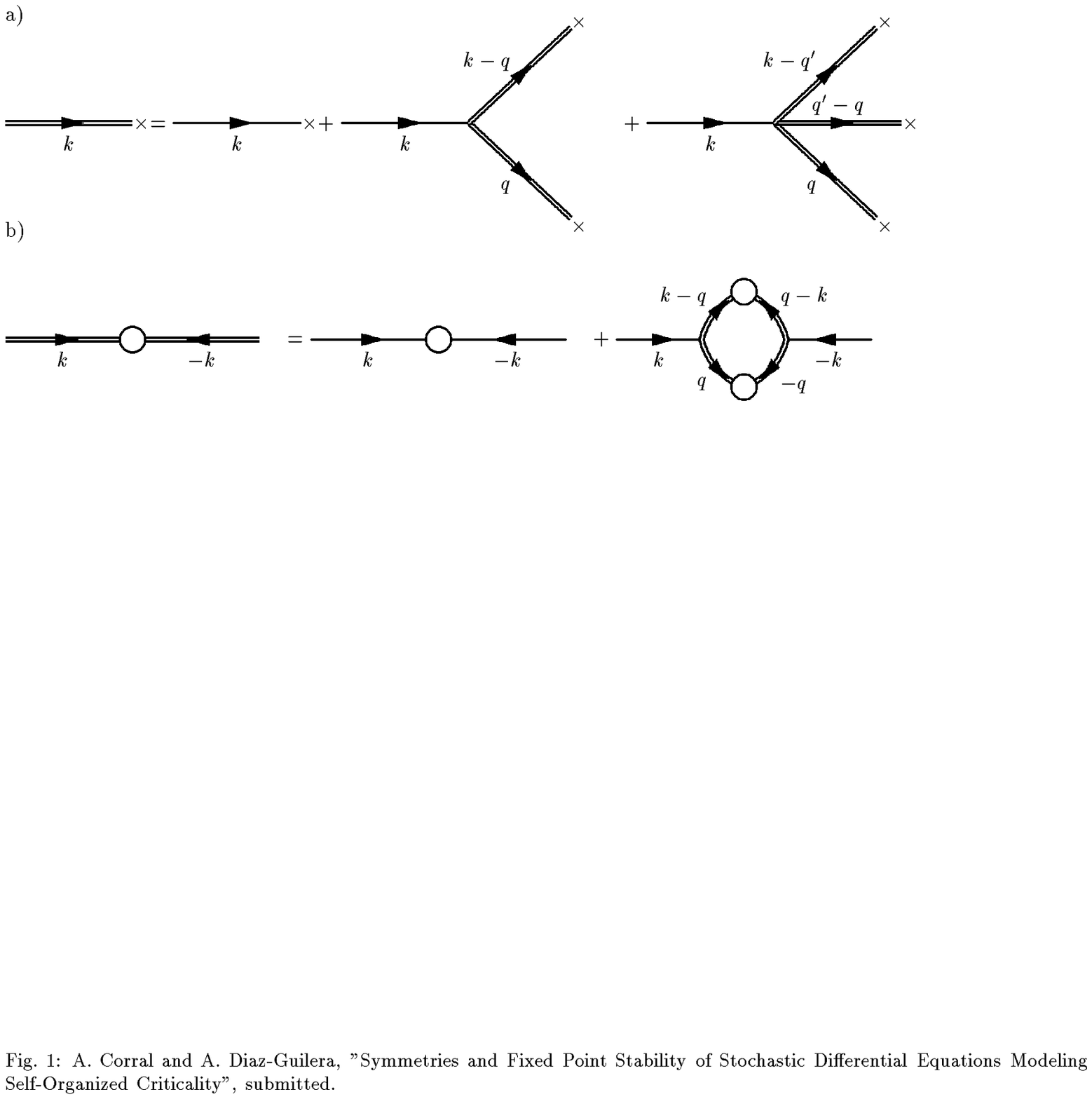}  
\label{diagram1} 
\caption{
Diagrammatic expressions for Eqs.  
(\protect{\ref{fourier_equation}}) and  
(\protect{\ref{correlationE}}), defined in the range $0<k<\Lambda$. 
The double bar with the cross $\times$ at its end is the order
parameter $E$, the single bar with the cross represents $G_0\eta$,
whereas the single bar alone is $G_0$.
A vertex with $n$ branches ($n=2$ or $3$ in the figure)
represents a convolution product of $n$ elements,
including a prefactor $-\lambda_n k^2 /(2 \pi)^{(n-1)(d+1)}$.
The circles correspond to the average over the noise.
} 
\end{figure} 
\begin{figure} 
[htbp] 
\epsfxsize=5truein 
\hskip 0.15truein
\epsffile{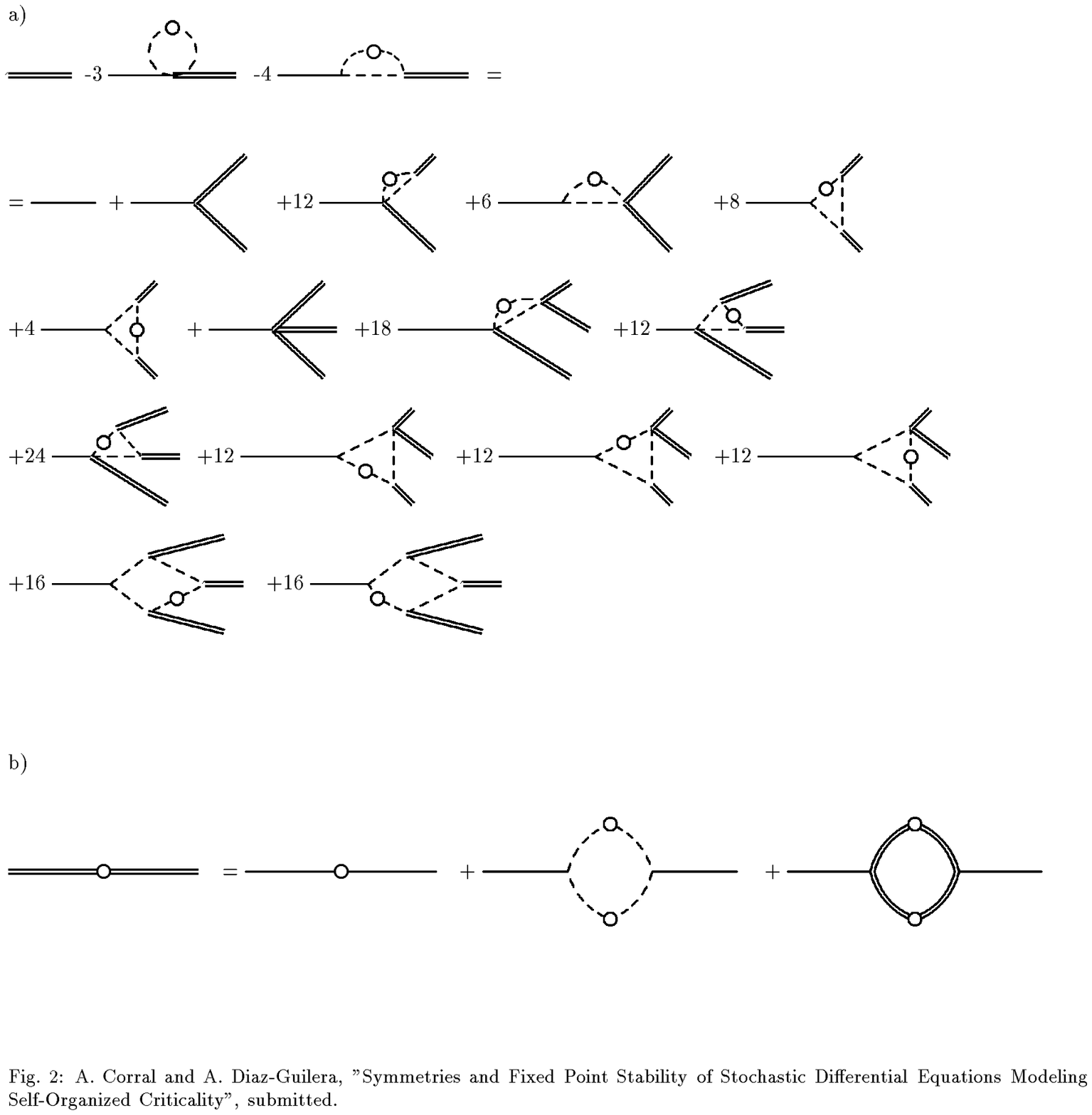}  
\label{diagram2} 
\caption{
Diagrams obtained after the first step of the DRG  
transformation. Now continuous lines correspond to the inner shell, 
whereas dashed lines correspond to the outer shell. The comparison 
with Fig. 1 
allows to define new coefficients. 
Observe that the new averages only affect the outer shell.
The notation has been simplified in respect to Fig. 1,
suppressing the symbol $\times$ at the end of the vertices
and also the arrows.
} 
\end{figure} 

The first step of the DRG transformation consists in splitting 
the Fourier space in two shells: an inner shell, that contains 
the slow modes, i.e., $0<k<e^{-l}\Lambda$, and an outer shell, 
containing the fast 
modes,  $e^{-l}\Lambda<k<\Lambda$. 
Both modes are coupled through the convolution products
in (\ref{fourier_equation}) and (\ref{correlationE}).  
We consider the diagrams for the slow modes and perform a  
perturbative expansion of the fast ones up to the lowest order
in the intensity of the noise (see Appendix for more details).  
Then we integrate out 
these modes by an average over the noise in the outer shell. 
After this transformation the resultant equations 
are shown diagrammatically in Fig. 2.  
It is clear that we can obtain new equations  
which are formally equivalent to the initial 
ones, (\ref{fourier_equation}) and (\ref{correlationE}), 
defining the new coefficients as the original ones plus the  
corresponding 
integrals over the outer shell. With the noise correlation 
(\ref{fourier_noise}) these integrals can be easily computed in the  
hydrodynamic limit ($\vec{k} \rightarrow 0$, $\omega \rightarrow 0$), 
as it is shown in the Appendix, 
and then the coefficients transform according to 
\begin{mathletters} 
\label{1st.step} 
\begin{equation} 
      \Gamma \rightarrow \Gamma, 
\label{1st.step.a} 
\end{equation} 
\begin{equation} 
       D \rightarrow D \left[1+3 \frac{I_d \Gamma  
      \lambda _3}{D^3}- 
       4\frac{I_d  \Gamma \lambda _2^2}{D^4} 
      \right], 
\label{1st.step.b} 
\end{equation} 
\begin{equation} 
      \lambda_2 \rightarrow \lambda _2 \left[  1 
      -18\frac{I_d \Gamma \lambda _3}{D^3} 
      +12\frac{I_d \Gamma  \lambda _2^2}{D^4}  
      \right], 
\label{1st.step.c} 
\end{equation} 
\begin{equation} 
      \lambda_3 \rightarrow\lambda _3 \left[ 1-18 
      \frac{I_d \Gamma  \lambda _3}{D^3}  
      +72\frac{I_d \Gamma   \lambda _2^2}{D^4} 
      -32\frac{I_d \Gamma \lambda _2^4}{D^5 \lambda _3} 
      \right], 
\label{1st.step.d} 
\end{equation} 
\end{mathletters}%
where  
\begin{equation} 
    I_d(l)=\frac{2 S_d}{(2 \pi)^d} \frac{1-e^{-l(d-4)}}{d-4} 
    \Lambda^{d-4},
\label{Id} 
\end{equation} 
and $S_d$ is the complete solid angle in $d$ dimensions. 
However, the new equations are only defined in the inner shell 
$0<k<e^{-l}\Lambda$.  
The second step allows to recover the original Brillouin 
zone $0<k<\Lambda$ by rescaling the equations using transformation  
(\ref{scale}), which in Fourier space writes 
\begin{equation} 
      \vec{k'} = e^{l} \vec{k},\hspace{3em} 
      \omega' = e^{z l}\omega,\hspace{3em} 
       E' = e^{ -(\chi +z+d)l }E. 
\label{fourierescale} 
\end{equation} 
The combined effect of both transformations, in the limit 
$l \rightarrow 0$, constitutes an infinitesimal DRG transformation, 
which gives the flow equations of the parameters
in parameter space. 
In these flow equations instead of $\lambda_2$ and $\lambda_3$ 
it is suitable to use  
the dimensionless coupling constants $\bar{\lambda}_2$ and  
$\bar{\lambda}_3$, given by 
\begin {equation} 
       \bar{ \lambda }_2^2=\frac{I_d^{(1)} \Gamma  \lambda _2^2}{D^4}  
       \hspace{2em} \mbox{and} \hspace{2em} 
       \bar{ \lambda }_3=\frac{I_d^{(1)} \Gamma \lambda _3}{D^3},
\label{dimensionless}
\end {equation} 
where $I_d^{(1)}= \left(dI_d/dl\right)_{l=0}= 
[2 S_d/(2 \pi)^d]\Lambda^{d-4}$. Then 
\begin{mathletters} 
\begin{equation} 
      \frac{d \Gamma }{dl}= \Gamma  \left[ 2z-2 \chi -d \right], 
\end{equation} 
\begin{equation} 
      \frac{dD}{dl}= 
       D \left[ z-2 
      -4\bar{\lambda}_2 ^2 
      +3 \bar{\lambda}_3 
      \right], 
\end{equation} 
\begin{equation} 
       \frac{d\bar{ \lambda }_2}{dl}= 
       \bar{ \lambda }_2 \left[ \frac{4-d}{2}+ 
       20 \bar{ \lambda }_2^2 -24\bar{ \lambda }_3 
       \right], 
\label{dl2edl} 
\end{equation} 
\begin{equation} 
      \frac{d\bar{ \lambda }_3}{dl}= 
      \bar{ \lambda }_3 \left[ 4-d 
      + 84\bar{ \lambda }_2^2 
      -27 \bar{\lambda }_3  
      -32 \frac{\bar{ \lambda }_2^4}{\bar{ \lambda }_3}\right]. 
\label{dl3edl} 
\end{equation} 
\label{flow} 
\end{mathletters}%
We are interested in the invariance of the parameters under DRG 
transformations. This means that we have to look for the fixed 
points of the flow equations; 
if we write Eqs. (\ref{flow}) as 
$d \alpha_i/d l=g_i(...\alpha_j...)$, 
where $\alpha_j$ represents any coefficient, then the fixed points 
verify  
$  
g_i(...\alpha_j^*...) = 0$, $\forall i 
$. 
Considering $D \ne 0$ and $\Gamma \ne 0$ we obtain 
four algebraic equations with four unknowns, $\chi$, $z$, 
$\bar{\lambda}_2$, and $\bar{\lambda}_3$; 
their solutions will give us the fixed points of the transformation,
$\bar{\lambda}_2^*$ and $\bar{\lambda}_3^*$,
as well as the values of the exponents $z$  and $\chi$ 
that guarantee that the DRG transformation leads to a
scale-free behavior.
Notice that the particular values of $\Gamma$ and $D$
play no role in the existence and location of the fixed points.
We can also find the stability 
of the fixed points under small perturbations using a linear stability analysis: 
the fixed point 
$\{\alpha_j^*\}$ is stable (i.e., an attractor)  
if all the eigenvalues (or their real parts) of the matrix 
$\partial{g_i}/\partial{\alpha_j}$ 
evaluated at this fixed point are negative. 
 
The results are the following: for $d>4$ one 
obtains six different fixed points, but the only stable one  
corresponds to 
\begin{equation} 
      \chi=\frac{\epsilon}{2},\hspace{3em} 
       z=2,\hspace{3em} 
      \bar{\lambda}_2^*=\bar{\lambda}_3^*=0, 
\end{equation} 
where as usual $\epsilon $ is defined as $\epsilon = d_c-d=4-d$. 
This is the trivial or Gaussian fixed point, that gives a  
normal (or Brownian) diffusive 
behavior because of the vanishing of the coupling constants. 
The values of the exponents do not correspond with those of the  
Edwards-Wilkinson model, used in the study of surface growth,  
because the noise correlation is different \cite{EW}. 
For $d<4$ this fixed point becomes unstable and the only stable one is 
\begin{equation} 
      \chi=\frac{7}{18}\epsilon,\hspace{3em} 
       z=2-\frac{\epsilon}{9},\hspace{3em} 
      \bar{\lambda}_2^*=0,\hspace{3em} 
      \bar{\lambda}_3^*=\frac{\epsilon}{27}, 
\label{fixed_point_<} 
\end{equation} 
that was unstable for $d>4$. 
In this case the diffusion is anomalous, to be more precise,  
the fact that $z<2$ gives a superdiffusive behavior 
in the hydrodynamic limit.  
Note that the one-loop expansion in the intensity of the noise 
$\Gamma$ gives a nontrivial fixed point which is expressed as a  
perturbation of the Gaussian one in a first order  
$\epsilon$-expansion. 
Observe also that the breaking of symmetry does not modify
the value of the fixed point obtained without taking into account
the even coupling constant $\lambda_2$ \cite{el26.177}.
Moreover, the fact that the nontrivial 
fixed point is an attractor of the dynamics contrasts with 
equilibrium critical phenomena, where this point is stable only 
along one direction. In this fact lies the difference between 
fine tuning of parameters for equilibrium systems at the 
critical point and self-organization towards criticality for 
nonequilibrium processes.   
 
Now we know the attractors in the parameter space,  
but this is not enough in our case; 
since our stochastic equation (\ref{series})
is derived directly from the discrete rules
of the BTW and Zhang's models,
we also need to know the basins of attraction of the
stable fixed points and 
whether our initial conditions, that is,  
the initial values of the coefficients 
corresponding to our physical problem, are inside these basins.  
These values for the dimensionless coupling constants (\ref{dimensionless})
can be calculated from Eqs. (\ref{D}) and  
(\ref{lambdan}), and they are 
\begin{equation} 
       (\bar{\lambda}^{o}_{2})^{2}= 
       \frac{1}{4} \frac{I_d^{(1)} \Gamma^{o}}{\alpha^2 E_c^2} 
       \frac{f^{(2)}(K)^2}{f^{(1)}(K)^4}, 
       \hspace{3em} 
       \bar{\lambda}_3^{o}= 
       \frac{1}{6} \frac{I_d^{(1)} \Gamma^{o}}{\alpha^2 E_c^2} 
       \frac{f^{(3)}(K)}{f^{(1)}(K)^3}, 
\label{initiallambda}        
\end{equation} 
result that also holds for $K=0$, where we obtain $\bar{\lambda}^{o}_{2}=0$
even for the Zhang's model. 
The superscript ''$o$'' indicates the initial value of the  
coefficient, that is, its value before any renormalization. 
As we have no restriction for $\Gamma^{o}$ 
(except that it has to be small), 
and $K$ can take any arbitrary real value, this implies that the initial 
dimensionless coupling constants will be defined in the following 
region: 
\begin{equation} 
       \bar{\lambda}_3^{o} < \frac{2}{3}(\bar{\lambda}^{o}_2)^2, 
\label{initialcond} 
\end{equation} 
having used for $f(x)$  
the explicit form given by Eq. (\ref{error}).  
 
Clearly, the stable fixed point for $d<4$ (\ref{fixed_point_<})
is outside the region of 
initial conditions defined by Eq. (\ref{initialcond}). 
It will be of the maximum interest
however to know whether these conditions will 
drive the system towards the nontrivial fixed point or not. 
We first consider $K=0$, which implies $\bar{\lambda}_2^{o}=0$,  
corresponding to the case studied in Ref. \cite{el26.177}. 
For $d<4$ one gets a different behavior depending on the sign 
of $\bar{ \lambda }_3^{o}$.  
Fig. 3 shows that when $\bar{ \lambda }_3^{o}$ 
is positive it flows towards the stable 
fixed point $\bar{ \lambda }_3^*=\epsilon /27$ giving a dynamical 
exponent $z=2-\epsilon /9$. 
A negative $\bar{ \lambda }_3^{o}$, 
which is our case of physical interest, flows away.  
An exact solution of Eq. (\ref{dl3edl}) with $\bar {\lambda}_2=0$ 
gives that $\bar {\lambda}_3 $ would reach $-\infty$ in a finite $l$ 
and then would reappear as $\bar {\lambda}_3 =\infty$, 
being then under the attraction of the nontrivial fixed point. 
However, our one-loop calculation forces the 
flow of the coupling constant along the parameter space
to stay of order $ \epsilon $, and one cannot sustain the validity 
of the preceding description. 
Then it is not possible to predict the  
renormalization of $\bar{\lambda}_3$. It will be either renormalized 
to $\bar{\lambda}_3^*$ or other fixed points 
will appear along the flow
(corresponding to strong coupling and 
not given by the one-loop $\epsilon$-expansion).
Therefore, the fixed point given by Eq. (\ref{fixed_point_<}), 
although it is an attractor, is unreachable  
from our initial conditions ($\bar{\lambda}_3^{o}<0$). 
For that reason the conclusions of Ref. \cite{el26.177} were incomplete.
On the other hand, above the upper critical dimension the system 
evolves towards the trivial fixed point $\bar{ \lambda }_3^*=0$  
giving a diffusive behavior with $z=2$ 
provided that $\bar{ \lambda }_3^{o}$ is not too much negative (see 
Fig. 3). 
This behavior of the fixed point $\bar{ \lambda }_3^*$ 
as a function of $\epsilon$ 
corresponds to a transcritical bifurcation. 
 
\begin{figure} 
[htbp] 
\epsfxsize=5truein 
\hskip 0.15truein
\epsffile{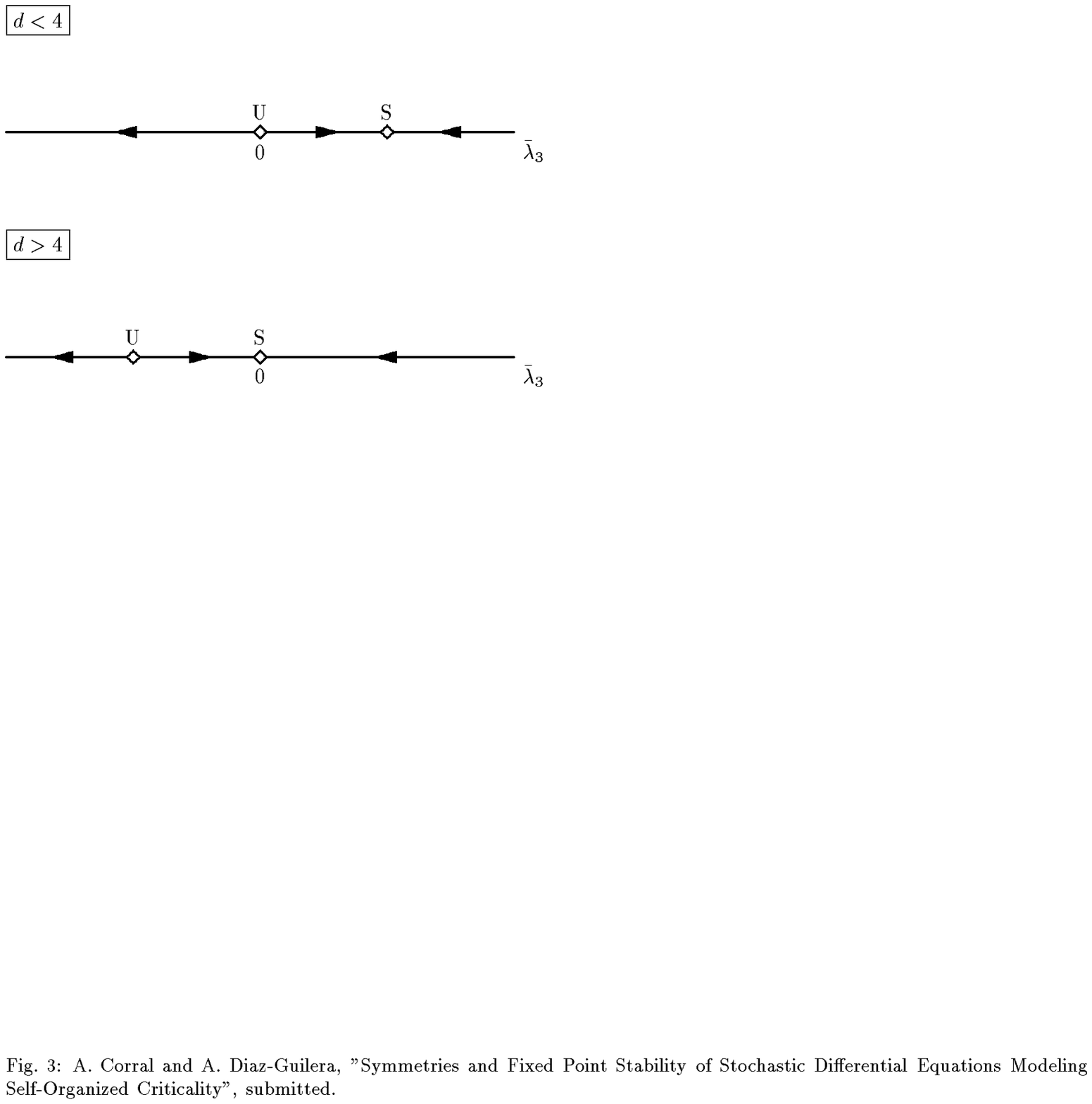}  
\label{flow3} 
\caption{Flow in $\bar{\lambda}_3$ space when only this 
nonlinear term is taken into account. The squares correspond to 
the stable (S) and unstable (U) fixed points and the arrows show 
the flow under DRG transformations.} 
\end{figure}

Now, by introducing the alternative regularization ($K \ne 0$), 
we will see the effect of the symmetry breaking. 
First of all, we insist that the stable fixed points 
are the same as for $K=0$, due to the fact that $\bar{ \lambda }_2$
renormalizes to zero.
Moreover, as can be seen in Fig. 4, 
where we have plotted the flow lines of Eq. (\ref{flow})  
obtained by numerical 
integration, the basin of attraction of the  
nontrivial fixed point is delimited by the parabola 
\begin{equation} 
      \bar{\lambda}_3=\frac{4}{7} \bar{\lambda}_2^2, 
\label{basin} 
\end{equation} 
which is also a particular solution of the flow equations,
no matter the value of $d$.
This parabola is inside the region defined by Eq. (\ref{initialcond}), 
and this fact implies that the new regularization makes
possible to reach the attractor for $d<4$ 
starting in the region of physical meaning. 
Using  Eqs. (\ref{initiallambda}) and (\ref{basin}) together with 
(\ref{error}) 
one gets that the condition to converge towards the nontrivial 
fixed point is 
$K^2 > \frac{7}{2}$.  
Then, the parameter that breaks the symmetry 
in the regularization of the step function, which in principle was 
arbitrary, determines the behavior of the system in the hydrodynamic 
limit.  
 
\begin{figure} 
[htbp] 
\epsfxsize=5truein 
\hskip 0.15truein
\epsffile{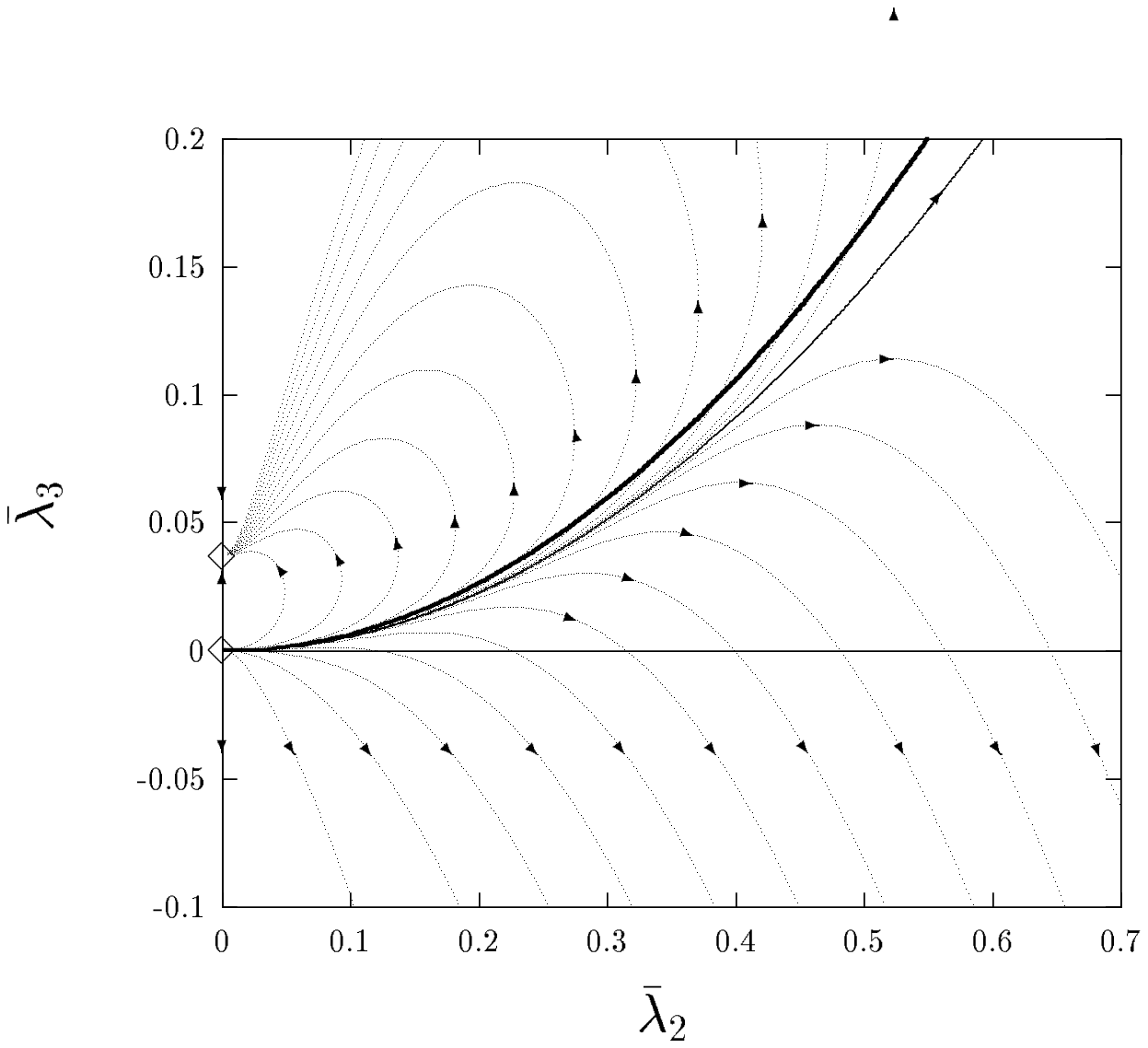}  
\label{flow23d3} 
\caption[]{Flow in ($\bar{\lambda}_2$, $\bar{\lambda}_3$) space  
for $d=3$ when both nonlinear terms are taken into account. 
In general for any $d<4$ the results are qualitatively the same. 
Dots correspond to the numerical integration of Eqs.  
(\ref{dl2edl}) and (\ref{dl3edl}), and the thin line is Eq. 
(\ref{basin}), that clearly delimits the basin of attraction of 
the nontrivial fixed point, as it is seen in the plot. 
Below the continuous thick line the values 
of the parameters correspond to our physical situation, Eq. 
(\ref{initialcond}). Squares correspond to the fixed 
points.
Observe that for $\bar{\lambda}_2=0$ we obtain the same
results as in Fig. 3.
} 
\end{figure} 
 
For $d>4$ the flow is more complex, because of the six fixed 
points, but the result is that convergence towards the Gaussian one 
also happens for our initial conditions, as Fig. 5 shows. 
The linear stability analysis of the fixed points gives the same  
results as the numerical integration shown in the figure. However 
this linear analysis fails for $d=4$, where all the fixed points 
collapse towards the Gaussian one. It is by means of the numerical 
integration that we verify that it is an attractor for the region 
above the parabola given by Eq. (\ref{initialcond}), but for the  
region below it is a repeller. This strange behavior appears because 
in $d=4$ we are at the bifurcation point. 
  
\begin{figure} 
[htbp] 
\epsfxsize=5truein 
\hskip 0.15truein
\epsffile{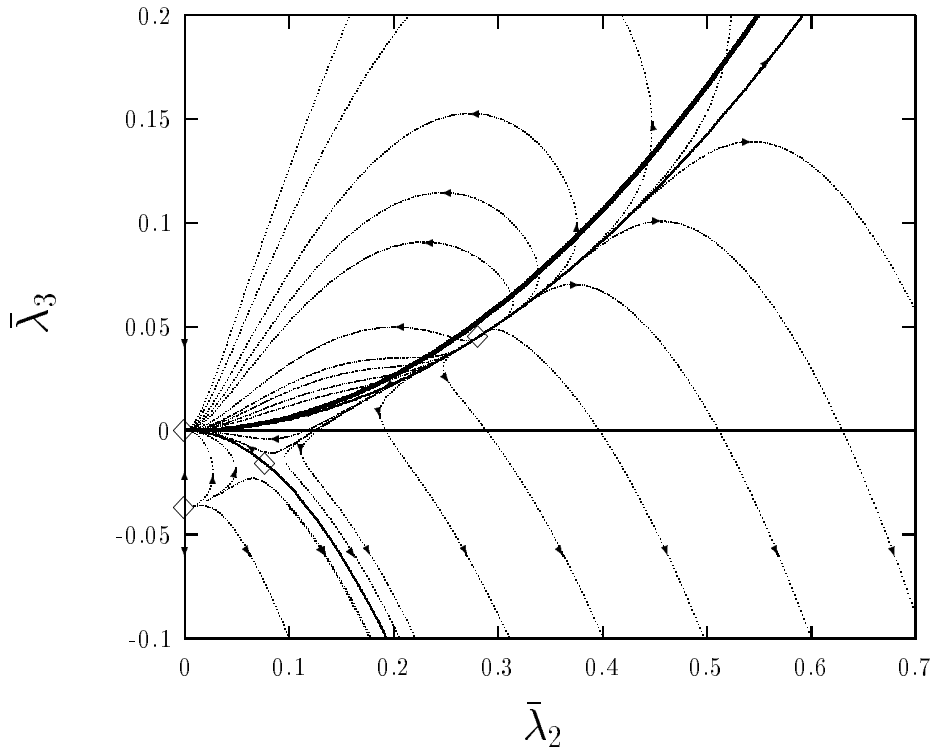}  
\label{flow23d5} 
\caption{Same as Fig. 4 but for $d=5$. The results 
hold for $d>4$. Only four of the six fixed points are shown 
because of the symmetry of the flow lines. In this case, the curve 
$\bar{\lambda}_3=-\frac{8}{3} \bar{\lambda}_2^2$, represented by  
another thin line, is the repulsive branch of the saddle 
point.} 
\end{figure} 
 
In Ref. \cite{el26.177} it was shown for the BTW model and $K=0$ 
that $\lambda_{2n}=0$, whereas for the Zhang's model, although 
the even coupling constants do not vanish, 
it was argued that their flow equations became decoupled
from the odd ones in the limit $\beta \rightarrow \infty$.
This fact enabled to establish the same universality class for  
both models, and to deal with only odd terms in Eq. 
(\ref{series}). Then, an expansion in the number of coupling 
constants  was
performed, whose extrapolation compares well with the 
results of the simulations \cite{prl63.470,pra45.8551}. 
Note that in the 
simulations one computes the dynamical exponent relating the 
characteristic length and lifetime of the avalanches, whereas 
within the DRG framework one computes the dynamical exponent 
from the fluctuations of the order parameter \cite{Giacometti}. 
The agreement between 
these calculations confirms the basic scaling hypothesis that 
in both cases length and time are related by
means of the same exponent. 
However, the problem of this calculation was that the nontrivial
fixed point was unreachable for the original equation.
 
In our approach, due to the symmetry breaking, we have to 
consider also the effect of even coupling constants.  
In the present work we have 
dealt with a restricted problem with only the lower-order even and 
odd coupling constants, $\lambda_2$ and $\lambda_3$, showing 
that $\lambda_2$ renormalizes to zero, supporting the 
calculation of Ref. \cite{el26.177}.  
Then the stable fixed points are not modified by the presence of an even
coupling constant in the model, but due to the symmetry breaking
that we have introduce, the nontrivial one is an attractor in the parameter
space when the parameters corresponding to the real model are taken
into account.
This behavior should be 
the same for any even coupling constant; actually,  
preliminary calculations including $\lambda_4$ and
$\lambda_5$ in Eq. (\ref{order3}) 
make us to suspect that all even coupling constants renormalize to 
zero.  This fact means that in the hydrodynamic limit the 
solution of both models has to be symmetric under parity 
transformations of the order parameter; then, for the BTW model 
the DRG restores the broken symmetry, whereas for the Zhang's 
model we conclude that its asymmetric nature is irrelevant in 
the behavior at large distances and long times. Therefore this 
validates the extrapolation performed in \cite{el26.177} since 
now we have showed that the symmetry breaking makes the stable 
fixed points reachable,
when starting in the region of physical interest in the space of 
parameters. 
Let us finally mention that in a recent work, Ghaffari and Jensen \cite{peyman}
perform a different extrapolation
of the same results which show a better agreement with large-scale
simulations and with real-space renormalization calculations
for the dynamical exponent \cite{Pietronero}.
It is noticeable than the same technique 
has been applied to the study of the effect
of dissipation in a uniformly driven BTW model \cite{peyman_2}.
 
\section{Conclusions} 
 
We have studied analytically two models which show self-organized 
criticality. 
The difference between them is that the second one (BTW) is symmetric 
under a parity transformation, 
whereas the first (Zhang's) model is not. 
From the microscopic 
rules one writes a effective long wave-length equation involving the 
threshold condition, which enters into the equation through a step 
function, making the equation unapproachable under this form. 
We have introduced a new regularization of the step function that 
breaks the symmetry of the BTW model. 
After a power series expansion, the equation is suitable  
for the application of the dynamic renormalization group,
although it contains an infinite number of relevant coupling constants.
In consequence one has to truncate at some point the expansion
in the coupling constants.
The results only have sense if it is possible to extrapolate
the values of the exponents up to an infinite number of coupling
constants.
We obtain the fixed points of the transformation in parameter 
space and study carefully their stability and basins of attraction. 
Then we find that with this regularization it is possible to reach the 
nontrivial fixed point for $d<4$, that was unreachable 
in a previous work, where symmetry was not broken. 
This means that in the hydrodynamic 
limit the models display ''scale invariance''. 
Moreover, in this limit we obtain a symmetric behavior under parity 
transformations for both models, and therefore the recovery of the  
broken symmetry for the BTW model 
and the irrelevance of this symmetry for Zhang's one. 
Although we have dealt with a simplified version of the problem,
we expect this behavior to be the same for the complete problem,
in the sense that all even coupling constants renormalize to zero,
validating the calculation of Ref. \cite{el26.177}.
The application of this technique should also be useful for 
other kinds of problems in which one deals with thresholds
or with an infinite number of nonlinear terms, 
for instance interface dynamics.
Moreover, the performed DRG calculation is interesting
because it provides an example showing how much important is
to know not only the stable fixed points of a DRG transformation
but also their basins of attraction.
It is remarkable the fact that a simple symmetry breaking
can solve the problem of the inaccessibility of the attractors
in parameter space.

The fact that the parameter that breaks the symmetry determines the  
behavior in the hydrodynamic limit is difficult to understand  
and we believe that it is an artificiality 
introduced in the calculation by the truncation in the  
coupling-constants expansion. 
We expect that higher orders in this expansion will give a behavior 
independent on the $K$ value. 
 
\section*{Acknowledgements} 
 
The authors wish to acknowledge A. Arenas, R. Cuerno, and 
C. J. P\'{e}rez for very fruitful discussions. 
A.C. has to thank the Spanish Ministerio de Educaci\'on y
Cultura for a scholarship.  
This work has been supported by CICyT of the Spanish 
Government, grants \#PB92-0863 and \#PB94-0897. 
 
\appendix\section*{} 
 
Here we present further details about the derivation of Eqs. 
(\ref{1st.step}), which give the transformation of the parameters 
after the first step of the DRG. 
Our starting points are Eqs. (\ref{fourier_equation}) and  
(\ref{correlationE}), i.e., the equations 
for $E(\vec k,\omega)$ and $<E(\vec k,\omega)E(\vec k',\omega')>$. 
As we have already mentioned, these equations are only defined for 
$0<k<\Lambda$.  
The DRG procedure consists of splitting   
the momentum space into two shells, an inner one, 
with $0<k<\Lambda e^{-l}$, and an outer one, with  
$\Lambda e^{-l}<k<\Lambda$. 
Then the magnitudes that depend on $\vec k$, like the energy $E$, 
split in the following way 
\begin{equation} 
    E(\vec k,\omega)=E^<(\vec k,\omega)+E^>(\vec k,\omega) 
    =E(\vec k,\omega)\Theta(\Lambda e^{-l}-k) 
    +E(\vec k,\omega)\Theta(k-\Lambda e^{-l}) 
\end{equation} 
where $\Theta(x)$ is again the Heaviside step function. 
This equality defines $E^<(\vec k,\omega)$ as the corresponding 
part of the energy in the inner shell, whereas $E^>(\vec k,\omega)$ 
is the same but defined in the outer shell. 
This separation also holds for the bare propagator $G_0$  
and the noise $\eta$. 
 
The DRG procedure eliminates the modes of the outer shell,  
within the same philosophy than the Kadanoff transformation  
does in real space. 
Then one is only interested in $E^<(\vec k,\omega)$ and  
$<E^<(\vec k,\omega)E^<(\vec k',\omega')>$, whose equations 
turn out to be equivalent to (\ref{fourier_equation}) and  
(\ref{correlationE}), but with additional 
terms due to the coupling between the two shells, via the convolution 
products. 
The fact that $E^>(\vec k,\omega)$ appears in the inner-shell  
equations allows a perturbative expansion in the form 
$E^>(\vec k,\omega)=G_0^>(\vec k,\omega)\eta^>(\vec k,\omega)+...$ 
(using the equivalent of Eq. (\ref{fourier_equation})  
but in the outer shell). 
Then the noise in the outer shell enters into the equation for 
$E^<(\vec k,\omega)$. 
A similar perturbative expansion is done for $<E^<(\vec k,\omega) 
E^<(\vec k',\omega')>$. 
By averaging over $\eta^>(\vec k,\omega)$, the contribution of the 
fast modes is eliminated from the inner shell.  
This is done up to one-loop order in the perturbative expansion, 
that is, the lowest order in the intensity of the noise $\Gamma$, 
which implies that it has to be small enough. 
This tedious calculation becomes more appealing using the diagrams 
of Fig. 1, instead of the corresponding equations. 
After this process, the relevant diagrams that survive the averaging  
are shown in Fig. 2. 
 
\begin{figure} 
[htbp] 
\epsfxsize=5truein 
\hskip 0.15truein
\epsffile{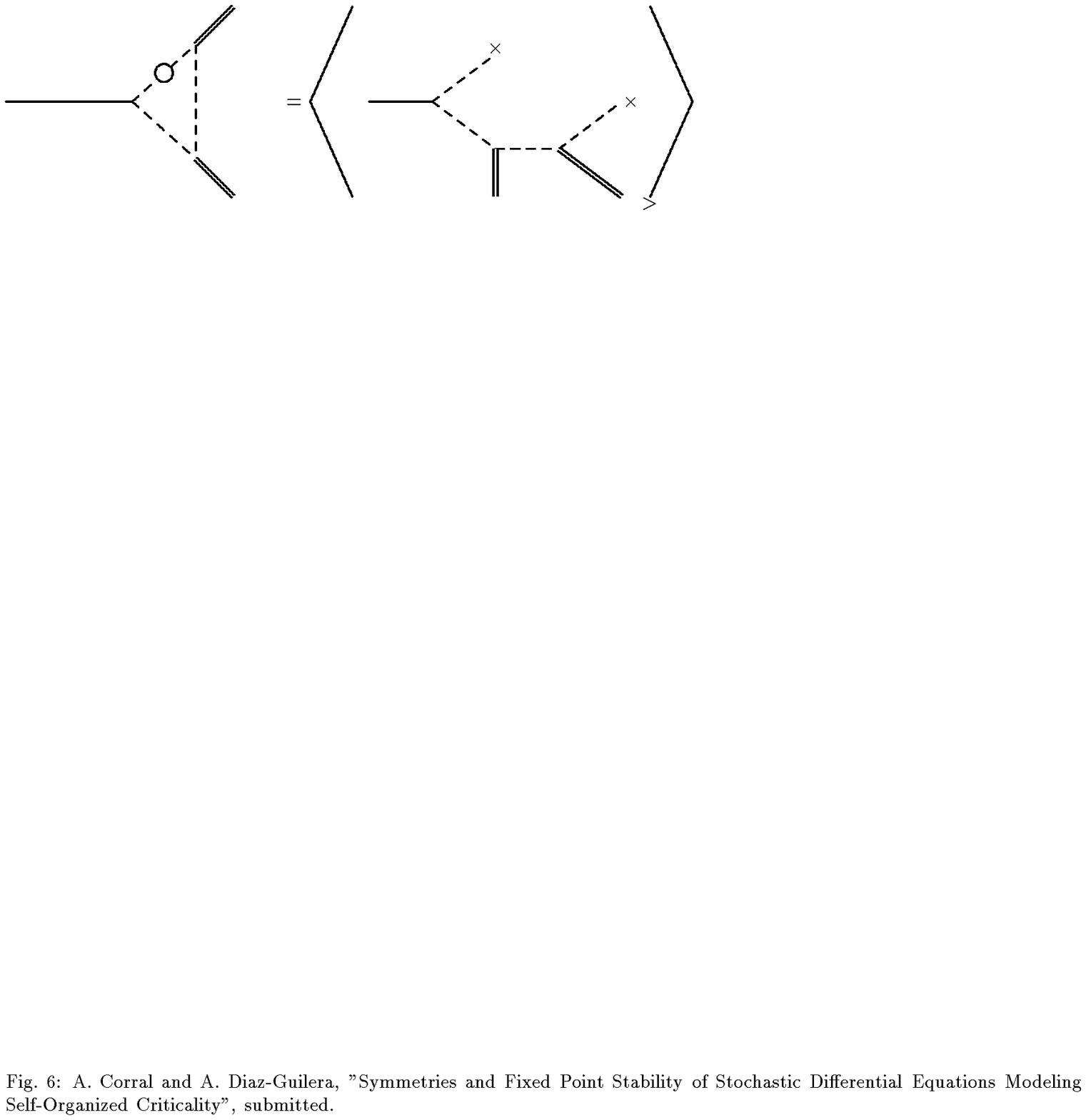}  
\label{exemple} 
\caption{Diagram computed in the Appendix as an example. The 
brackets stand for an average over the outer shell.} 
\end{figure} 
 
As an example let us consider one of them, shown 
in Fig. 6 and denoted by $V(\vec k,\omega)$: 
\begin{eqnarray} 
\nonumber 
      V(\vec k,\omega)= 
      <\frac{-\lambda_2 k^2}{(2 \pi)^{d+1}}G_0^<(\vec k,\omega) 
      \int d^dq \ d\Omega G_0^>(\vec k-\vec q,\omega-\Omega) 
      G_0^>(\vec q,\Omega)\eta^>(\vec q,\Omega)\times 
\\ 
\nonumber 
      \times \frac{-\lambda_2 (\vec k-\vec q)^2}{(2 \pi)^{d+1}} 
      \int d^dq' \ d\Omega' G_0^>(\vec k-\vec q-\vec {q'}, 
      \omega-\Omega-\Omega')E^<(\vec{q'},\Omega')\times 
\\ 
\nonumber 
      \times \frac{-\lambda_2 (\vec k-\vec q-\vec {q'})^2} 
      {(2 \pi)^{d+1}} 
      \int d^dq''\ d\Omega'' G_0^>(\vec k-\vec q-\vec {q'}-\vec {q''}, 
      \omega-\Omega-\Omega'-\Omega'')\times 
\\ 
      \times\eta^>(\vec k-\vec q-\vec {q'}-\vec {q''}, 
      \omega-\Omega-\Omega'-\Omega'')E^<(\vec{q''},\Omega'')>_>, 
\end{eqnarray} 
where the symbol $<...>_>$ stands for an average over the outer shell. 
Using the noise correlation given by Eq. (\ref{fourier_noise}) we can  
integrate over $\Omega$, $\Omega''$, and $\vec{q''}$, and then we have 
\begin{eqnarray} 
\nonumber 
     V(\vec k,\omega)= 
     -\frac{2\lambda_2^3 \Gamma}{(2 \pi)^{2d+1}}k^2  
     G_0^<(\vec k,\omega) 
     \int d^dq' \ d\Omega'  
     E^<(\vec{q'},\Omega')E^<(\vec k -\vec {q'},\omega-\Omega')\times 
\\     
     \times\int d^dq \ G_0^>(\vec k-\vec q,\omega)G_0^>(\vec q,0) 
     (\vec k-\vec q)^2 G_0^>(\vec k-\vec q-\vec {q'},\omega-\Omega') 
     (\vec k-\vec q-\vec {q'})^2 G_0^>(\vec q,0). 
\label{fig6} 
\end{eqnarray} 
As the bare propagator is a known function, given by Eq. 
(\ref{propagator}), we are also able to perform the integral over  
$\vec{q}$, that is 
\begin{equation} 
\nonumber 
    \int d^d q \ [...]=\int d^dq \ G_0^>(\vec k-\vec q,\omega) 
    {G_0^>}^2(\vec q,0)(\vec k-\vec q)^2  
    G_0^>(\vec k-\vec q-\vec {q'},\omega-\Omega') 
    (\vec k-\vec q-\vec {q'})^2 . 
\end{equation} 
This integral is a function of $\vec k,\omega,\vec {q'}$, and 
$\Omega'$. However, we are going to evaluate it in the hydrodynamic  
limit, by taking $\vec k,\vec {q'} \rightarrow 0$ and $\omega,\Omega' 
\rightarrow 0$. Then 
\begin{equation} 
\nonumber 
   \int d^d q \ [...]=\int d^dq \ q^4 {G_0^>}^2(-\vec q,0) 
   {G_0^>}^2(\vec q,0)= 
   \frac{S_d}{D^4}\int_{\Lambda e^{-l}}^{\Lambda}q^{d-5}dq= 
   \frac{S_d}{D^4}\frac{\Lambda^{d-4}}{d-4}\left(1-e^{-l(d-4)}\right). 
\end{equation} 
It is easy to check that this result is also valid for $d=4$. 
We have used the explicit form of the bare propagator  
(\ref{propagator}) and also that $d^dq\ =S_d q^{d-1}dq$, 
with $S_d$ the complete solid angle in $d$ dimensions, that is, 
the area of a unit ($d$)-sphere.  
Then, by making use of Eq. (\ref{Id}), we obtain 
\begin{equation} 
    \int d^d q \ [...]=\frac{(2 \pi)^d I_d(l)}{2 D^4}, 
\nonumber 
\end{equation} 
and substituting into Eq. (\ref{fig6}), 
\begin{equation} 
      V(\vec k \rightarrow 0,\omega \rightarrow 0) \longrightarrow 
     -\frac{\lambda_2 k^2}{(2\pi)^{d+1}} 
       G_0^<(\vec k,\omega) (E^<*E^<)(\vec k,\omega) 
       I_d(l)\frac{\lambda_2^2 \Gamma}{D^4}. 
\label{fig6.2} 
\end{equation} 

It is clear from Fig. 2 that after the first step of the DRG we 
have the  
same as at the beginning (Fig. 1), but defined only in the inner  
shell, plus a lot of diagrams of the same type as the one in Fig. 6. 
These diagrams, which contain integrals over the outer shell,  
renormalize the other diagrams that are only defined in the inner  
shell. For instance, if we consider the following diagram 
%
%
%
\begin{equation}
\unitlength=1mm
\linethickness{0.4pt}
\begin{picture}(40.00,30.00)
\put(10.00,20.00){\line(1,0){20.00}}
\put(30.00,20.00){\line(1,1){10.00}}
\put(40.00,28.00){\line(-1,-1){8.00}}
\put(30.00,20.00){\line(1,-1){10.00}}
\put(40.00,12.00){\line(-1,1){8.00}}
\end{picture}
=       -\frac{\lambda_2 k^2}{(2\pi)^{d+1}} 
       G_0^<(\vec k,\omega) (E^<*E^<)(\vec k,\omega) 
\label{dibujo} 
\end{equation} 
and compare it with with Eq. (\ref{fig6.2}) we observe only an 
additional term $I_d(l)\lambda_2^2 \Gamma/D^4$ that comes from the  
outer shell integration. So, the diagram shown in Fig. 6 contributes 
to the renormalization of (\ref{dibujo}), that is, renormalizes the 
coupling constant $\lambda_2$. As Fig 2(a) states, the diagram in Fig. 
6 appears eight times in the perturbative expansion, and the new  
$\lambda_2$, after the first step of the transformation will be  
modified by 
\begin{equation} 
    \lambda_2 \rightarrow \lambda_2 
    \left(1+8I_d(l)\frac{\lambda_2^2\Gamma}{D^4}+...\right). 
\nonumber 
\end{equation} 

In the same way one can perform the outer-shell integrals of the rest 
of diagrams in Fig. 2(a). A general result for its contribution to the  
renormalization of any coupling constant $\lambda_n$ or to the  
diffusion coefficient $D$ (which will be referred here also as 
$-\lambda_1$) is given by 
\begin{equation} 
\label{outer} 
    (-1)^{v-1}I_d(l)\frac{\Gamma}{D^{v+1}} 
    \frac{\prod_{m=1}^{v} \lambda_{b(m)}}{\lambda_B} 
\end{equation} 
where $v$ is the number of vertices each diagram has, three for our  
example (since the dashed line in Fig. 6 forms a triangle), 
$b(m)$ is the number of branches of the $m$-th vertex, 
two for each one in the example, and $B$ is the number of branches 
of the diagram that is renormalized (two in the example), 
and fulfills $B=\sum_{m=1}^v b(m)-v-1$. 
Note that the magnitude in Eq. (\ref{outer}) is dimensionless. 
Using this equation and Fig. 2(a) the derivation of Eqs. 
(\ref{1st.step.b}-\ref{1st.step.d}) is then straightforward. 
 
For the renormalization of the intensity of the noise $\Gamma$ 
we have only one diagram, the dashed one in Fig. 2(b). 
It is immediate to see that the integral over the outer shell, 
no matter is value, is multiplied by a factor $k^2{k'}^2$. Then 
\begin{equation} 
\nonumber 
    \Gamma \rightarrow \Gamma 
    \left(1+A k^2{k'}^2 \frac{\lambda_2^2\Gamma}{D^4}+...\right), 
\end{equation} 
where $A$ is simply a numeric factor, 
and hence, in the hydrodynamic limit and up to one-loop order,  
the intensity of the noise is not renormalized, after the 
first step of the DRG, as Eq. (\ref{1st.step.a}) states. 
 


\end{document}